\documentclass[aps,prx,reprint,groupedaddress]{revtex4-1}

\usepackage{amsmath} 
\usepackage{amssymb}	
\usepackage{graphicx}
\usepackage{epstopdf}

\renewcommand{\vec}[1]{\mathbf{#1}}
\newcommand{\curl}[1]{\vec{\nabla}\times\vec{#1}}
\renewcommand{\div}[1]{\vec{\nabla}\cdot\vec{#1}}
\newcommand{\grad}{\vec{\nabla}}

\begin{document}


\title{Diffractive Interface Theory: \\Nonlocal polarizability approach to the optics of metasurfaces}


\author{Christopher M. Roberts}
\email[]{christopher\_roberts@student.uml.edu}
\author{Sandeep Inampudi}
\author{Viktor A. Podolskiy}
\email[]{viktor\_podolskiy@uml.edu}
\affiliation{Department of Physics and Applied Physics, University of Massachusetts Lowell, Lowell, MA 01854}


\date{\today}

\begin{abstract}
We present a formalism for understanding the electromagnetism of metasurfaces, optically thin composite films with engineered diffraction. The technique, diffractive interface theory (DIT), takes explicit advantage of the small optical thickness of a metasurface, eliminating the need for solving for light propagation inside the film and providing a direct link between the spatial profile of a metasurface and its diffractive properties. Predictions of DIT are compared with full-wave numerical solutions of Maxwell's equations, demonstrating DIT's validity and computational advantages for optically thin structures.  Applications of the DIT range from understanding of fundamentals of light-matter interaction in metasurfaces to efficient analysis of generalized refraction to  metasurface optimization. 
\end{abstract}


\maketitle
\section{Introduction}
Metasurfaces, optically thin structures with engineered diffraction, in the past few years have gained attention as a new platform for controlling the flow of light for ultra-compact photonics\cite{bomzon00, hasman04, hasman05, yu11,sun12,xiao13,ni12,farmahini13,alu11,shalaev13,zhang13,Zhao11}.  However, despite significant progress in fabrication and experimental characterization of metasurfaces, there is lack of analytical and numerical tools aimed at understanding and optimizing their optical behavior. As a result of the quasi-two-dimensional geometry of metasurfaces, the majority of light-matter interaction occurs in the near-field proximity of the interfaces.  At the same time, conventional techniques for numerical calculation of light interaction with composite systems (metamaterials) have been developed with volumetric materials in mind, \cite{moharam81,FEM,FDTD} and as result, typically are inefficient in understanding the optics of metasurfaces. Likewise, traditional effective medium theories\cite{milton2002,noginov2011}, which assign effective permittivities and permeabilities to volumetric structures cannot be used to describe the diffractive optics of metasurfaces that, in some sense, are more akin to the two-dimensional graphene than to the traditional volumetric-metamaterials.  Here we present a formalism that takes advantage of the small optical thickness of a metasurface and can successfully be used to predict its optical response. This technique can be used for development of analytical description of light-metasurface interaction and for rapid design and prototyping of metasurfaces for a wide range of optical applications. 

Metasurfaces are typically based on arrays of planar optical resonators, fabricated at the interface between two dielectrics, arranged to produce diffracted waves of pre-designed amplitude and direction. Previous studies have focused on the induced phase discontinuity at a metasurface. Generalized Snell's Law \cite{yu11,ni12} as well as standard diffraction theory\cite{larouche12} can be used to predict the {\it direction} of the diffracted beam. At the same time, calculation of the phase gradient or calculations of the {\it amplitudes} of the diffracted beam require full-wave solutions of Maxwell's equations. The vast majority of studies rely on finite element method (FEM)\cite{FEM} or finite-difference time domain (FDTD)\cite{FDTD} techniques to solve for light interaction with resonant grating. However, while possible, these solutions require meshing the relatively large (multiple wavelength) regions of space with deep subwavelength (order of grating element) resolution, leading to resource-intensive and time-consuming calculations. An alternative numerical technique, rigorous coupled wave analysis (RCWA)\cite{moharam81}, requires solving a large eigenvalue problem, once again making it impractical for design and optimization of real-life structures. More importantly, even when possible, calculation of field distribution is rarely sufficient for {\it understanding} the underlying physics which yields this distribution. Successful attempts to gain insight into basics of light interaction with elements of metasurfaces are limited to a few simple resonator shapes \cite{yu11,bomzon00,wong14,zhang13,Zhao11}. 

\begin{figure}[b]
\includegraphics[width=3in]{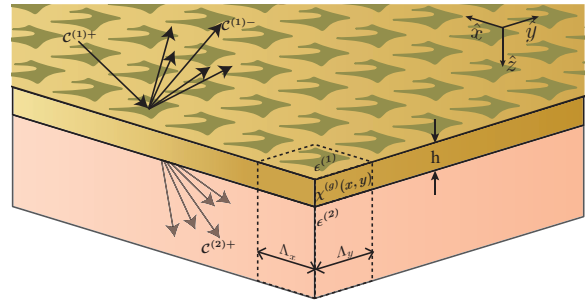}
\caption{Schematic geometry of metasurface structure; DIT approximation is valid in regime $h\ll\lambda_0$}\label{fig:basicGeo}
\end{figure}

In this work, we present a general formalism for understanding the link between the spatial distribution of polarizability of a metasurface and the resulting diffraction performance of the structure. In contrast to the majority of existing analytical and numerical techniques, the proposed approach approximates the metasurface as a thin polarizable sheet, and thus it takes explicit advantage of quasi-two-dimensional geometry. The spatial inhomogeneity of the polarization distribution results in nonlocal (wavevector-dependent) polarizability, which in turn results in multiple reflected and refracted (diffracted) beams\cite{pekar58,wells14}. Amplitudes of these beams can be related to the amplitude of the incident beam via generalized boundary conditions. Our technique mostly resembles the work that characterizes metasurfaces using generalized sheet transition currents \cite{holloway09}.  However, this previously developed method relies on a priori knowledge of the optical response of $0^{th}$ order diffraction, a major limitation for the design of complex metasurfaces, is addressed by our proposed formalism. In some sense, our formalism merges RCWA, Transfer Matrix Method (TMM)\cite{rytov56,yeh77}, and Effective Medium Theories (EMT), resulting in a general approach to optics of optically thin diffractive systems: diffractive interface theory (DIT). 

\section{General Formalism}

As mentioned above, we aim to derive the formalism for understanding reflection, refraction, and diffraction at an interface between two homogeneous materials covered with an optically thin inhomogeneous layer (Fig.\ref{fig:basicGeo}). We assume that all elements of the structure are non-magnetic and have a linear responses to electric fields. Specifically, this linear response yields a polarization field inside the material with $P=\chi E$ with $E$ being the excitation field and $\chi$ being polarizability. In bulk materials, polarizability is related to permittivity via 

\begin{equation}
\epsilon=1+4\pi\chi.\label{eq:eps}
\end{equation} 

Permittivity, in turn, can be used to calculate the dispersion of the waves propagating inside such a material, as well as amplitudes and directions of these waves when the material is illuminated by incident light etc. In 2D films, however, the introduction of bulk permittivity is not meaningful. Therefore, we represent distribution of polarizability across the structure as

\begin{equation}
\chi(x,y,z)=\chi^{(b)} + \chi^{(g)}(x,y)\delta(z)h.
\end{equation}
Here, $\chi^{(b)}$ represents polarizability of bulk components of the system, while $\chi^{(g)}$ describes polarizability of material constructing the diffractive (grating) part of the structure, and $h$ is the physical height of the grating layer. Note that both $\chi^{(b)}$ and $\chi^{(g)}$ correspond to volumetric polarizability of the materials comprising metasurface and, for implementation purposes, can be related to permittivity via Eq. (\ref{eq:eps}). Assuming a harmonic time-dependence $\vec{E},\vec{H}\propto e^{-i\omega t}$, and the absence of free charges ($\div{E}=-4\pi\div{P}$), Maxwell's equations can be represented as

\begin{align}
\curl{E}=&\frac{i\omega}{c}\vec{H}\label{eq:curlE}\\
\curl{H}=&-\frac{i\omega}{c}(\vec{E}+4\pi\vec{P})\label{eq:curlH}, 
\end{align}
resulting in 
\begin{equation}
4\pi\grad(\div{P})-\grad^2\vec{E}=\frac{\omega^2}{c^2}(\vec{E}+4\pi\vec{P})\label{eq:waveEq}.
\end{equation}

As shown, for example, in Ref.\cite{jackson}, the set of waves propagating inside the system (modes of the structure) are completely determined by bulk components of the system through bulk $[\chi^{(b)}]$ component of polarizability. The polarization sheet $\chi^{(g)}$ affects only boundary conditions. 

It can be explicitly shown that the tangential components of both magnetic and electric fields are often discontinuous across a metasurface of finite, but small, thickness. The discontinuities of the magnetic field can be straightforwardly calculated by integrating Eq.(\ref{eq:curlH}) over ``Amperian loops'' in the $xy$ and $yz$-planes, leading to 
\begin{align}
\Delta H_x &= -\frac{4\pi i \omega h}{c}\chi^{(g)}(x,y)E^{\rm avg}_y\nonumber\\
\Delta H_y &= \frac{4\pi i \omega h}{c}\chi^{(g)}(x,y)E^{\rm avg}_x\label{eq:dH}, 
\end{align}
where $E^{\rm avg}_\alpha(x,y)=[E_\alpha(x,y,z=+0)+E_\alpha(x,y,z=-0)]/2$. Similarly, integration of Eq.(\ref{eq:waveEq}) yields discontinuities for the tangential components of the electric field: 
\begin{align}
\Delta E_x=&-4\pi \frac{\partial P_z}{\partial x}\nonumber\\
\Delta E_y=&-4\pi \frac{\partial P_z}{\partial y}\label{eq:dE}
\end{align}
An alternate but equivalent derivation of these boundary conditions can be found in Ref. \cite{idemen88}

\section{DIT for Periodic Metasurfaces}

We now apply the generalized boundary conditions presented above to derive diffractive interface theory (DIT), the formalism relating the refractive and diffractive properties of the periodic metasurface to its geometry and composition. We assume that the metasurface is periodic in the $xy$ plane (Fig.\ref{fig:basicGeo}) and is surrounded by two homogeneous materials. For simplicity, we will characterize the quantities referring to the materials above, below, and inside the grating by indices $(b)=(1),(2)$, and $(g)$, respectively. We assume that the structure is excited by a plane-wave with in-plane component of wavevector $\vec{k_0}=\{k_{0x}\hat{x}+k_{0y}\hat{y}\}$. 

As with any periodic system, the Bloch theorem implies that electromagnetic field in the regions $(1)$ and $(2)$ can be represented as linear combination of (plane) waves with in-plane components of the wavevector separated by the multiple of the inverse lattice vector.  Explicitly, the $\alpha$ Cartesian components of the field in the region $(b)$ are written as

\begin{align}
E_\alpha^{(b)}(r)&=\sum_{j}\mathcal{E}_{\alpha,j}^{(b)}(z)e^{ik_{j,x}x+ik_{j,y}y}\nonumber\\
H_\alpha^{(b)}(r)&=\sum_{j}\mathcal{H}_{\alpha,j}^{(b)}(z)e^{ik_{j,x}x+ik_{j,y}y}\label{eq:Hamp}.
\end{align}

Here, we utilize a common diagonalization technique used in implementations of RCWA and PWEM\cite{PWEM} in which the index $j$ spans the domain of integers for one-dimensional gratings and the domain of integer ordered pairs for two-dimensional gratings. The in-plane components of the wavevector of the waves are independent in the bulk regions $(b)$ and are given by
\begin{eqnarray}
{k}_{j,x}=k_{0x}+\frac{2\pi j_x}{\Lambda_x}; k_{j,y}=k_{0y}+\frac{2\pi j_y}{\Lambda_y}
\end{eqnarray}
with the remaining component controlled by the dispersion of the bulk layers $k^{(b)^2}_{j,z}=\epsilon^{(b)}\omega^2/c^2-k_{j,x}^2-k_{j,y}^2$. 

It can be shown that out of six Cartesian components of the electromagnetic fields, only four components are independent; here we chose in-plane ($xy$) components of electric and magnetic fields as the independent fields which are used to enforce the boundary conditions [Eqs.(\ref{eq:dH},\ref{eq:dE})]. We further relate these four fields to the four sets of amplitudes of plane waves propagating in $+z$ and $-z$ direction with TE or TM polarization (see Appendix).

\begin{equation}
\left(\begin{array}{c} \mathcal{E}^{(b)}_{x} \\ \mathcal{E}^{(b)}_{y}  \\ \mathcal{H}^{(b)}_{x} \\ \mathcal{H}^{(b)}_{y} \end{array}\right)
= \mathbb{F}^{(b)} \left( \begin{array}{c} \mathcal{C}^{(b)+}_{TE} \\ \mathcal{C}^{(b)+}_{TM} \\ \mathcal{C}^{(b)-}_{TE} \\ \mathcal{C}^{(b)-}_{TM} \end{array}\right)\label{eq:fields}
\end{equation}
where $\mathcal{(E,H,C^{\pm})}$ are column-vectors with entries spanning the set of $j$ indices.

Therefore, the problem of solving Maxwell's equations in the space surrounding the metasurface is reduced to the problem of finding the amplitudes a plane-wave expansion of electromagnetic fields in the bulk layers surrounding the metasurface. 

The polarizability of the periodic metasurface can be expressed as 
\begin{equation}
\chi^{(g)}(x,y)=\sum_{j}\hat{\chi}^{(g)}_{j}e^{i(k_{j,x}x+k_{j,y}y)}
\end{equation}
The parameter $\hat{\chi}^{(g)}_j\equiv \hat{\chi}^{(g)}({k_{j,x},k_{j,y}})$ describes the effective polarizability of the metasurface. Its nonlocal nature\cite{pekar58,wells14} (dependence on the wavevector) reflects long-range correlation and leads to diffraction, and allows the coupling of modes with different wavevectors during the interaction of light with the metasurface.

After a direct substitution of Eq.(\ref{eq:fields}) into the Fourier transformations of Eqs.(\ref{eq:dH},\ref{eq:dE}) and some straightforward mathematical transformations (see Appendix), we obtain the following block-matrix relationship: 
\begin{widetext}
\begin{equation}
 \begin{pmatrix}
  \mathbb{I} & \mathbb{O} & \mathbb{K}_x \mathbb{X}_{z} \mathbb{K}_y & -\mathbb{K}_x \mathbb{X}_{z} \mathbb{K}_x \\
  \mathbb{O} & \mathbb{I} & \mathbb{K}_y\mathbb{X}_{z} \mathbb{K}_y & -\mathbb{K}_y \mathbb{X}_{z} \mathbb{K}_x \\
  \mathbb{O}  & \mathbb{X}_{xy}  & \mathbb{I} & \mathbb{O}  \\
  -\mathbb{X}_{xy}  & \mathbb{O} & \mathbb{O} &\mathbb{I} \\ 
\end{pmatrix}
\mathbb{F}^{(1)}
\left( \begin{array}{c} \mathcal{C}^{(1)+}_{TE} \\ \mathcal{C}^{(1)+}_{TM} \\ \mathcal{C}^{(1)-}_{TE} \\ \mathcal{C}^{(1)-}_{TM} \end{array}\right)=
 \begin{pmatrix}
  \mathbb{I} & \mathbb{O} & -\mathbb{K}_x \mathbb{X}_{z} \mathbb{K}_y & \mathbb{K}_x \mathbb{X}_{z} \mathbb{K}_x \\
  \mathbb{O} & \mathbb{I} & -\mathbb{K}_y\mathbb{X}_{z} \mathbb{K}_y & \mathbb{K}_y \mathbb{X}_{z} \mathbb{K}_x \\
  \mathbb{O}  & -\mathbb{X}_{xy}  & \mathbb{I} & \mathbb{O}  \\
  \mathbb{X}_{xy}  & \mathbb{O} & \mathbb{O} &\mathbb{I} \\ 
\end{pmatrix}
\mathbb{F}^{(2)}
\left( \begin{array}{c} \mathcal{C}^{(2)+}_{TE} \\ \mathcal{C}^{(2)+}_{TM} \\ \mathcal{C}^{(2)-}_{TE} \\ \mathcal{C}^{(2)-}_{TM} \end{array}\right)\label{eq:MG}
\end{equation}
\end{widetext}
where $\mathbb{I}$ represent identity matrices, $\mathbb{O}$ represent matrices filled with zeros, $\mathbb{X}_{xy}=-\frac{2\pi i \omega h}{c}\tilde{\chi}^{(g)}$ and $\mathbb{X}_z=-\frac{2\pi i c h}{\omega}\tilde{\chi}^{(g)}(\mathbb{I}+4\pi\tilde{\chi}^{(g)})^{-1}$, with the $mn$ element of matrix $\tilde{\chi}^{(g)}$ given by $\tilde{\chi}_{g_{mn}}=\hat{\chi}_{g_{m-n}}$, and $\mathbb{K}_\alpha$ being diagonal matrices with elements $\mathbb{K}_{\alpha_{mm}}=k_{m,\alpha}$.

Eq.(\ref{eq:MG}) represents the main result of this work. It essentially represents an extension of transfer matrix formalism\cite{rytov56,yeh77} to the domain of diffractive systems. As such, it can be used to 
\begin{itemize} 
\item calculate the dispersion of the guided modes of the system by solving for dependence of $\vec{k_0}(\omega)$ that produces the solution with  $\mathcal{C}^{(1)+}= \mathcal{C}^{(2)-}=0;  \mathcal{C}^{(1)-},\mathcal{C}^{(2)+}\neq 0$

\item calculate the diffraction (generalized refraction) of metasurfaces by computing amplitudes of diffracted fields $\mathcal{C}^{(1)-}$ and $\mathcal{C}^{(2)+}$ as a function of incident fields  $\mathcal{C}^{(1)+}$ and  $\mathcal{C}^{(2)-}$

\item design, optimize, or retrieve the parameters of the metasurface by calculating the matrix $\hat{\chi}^g$, and in the end, distribution $\chi^{(g)}(x,y)$ that realizes the desired or observed optical performance
\end{itemize}

We now illustrate these applications of the proposed formalism on several representative examples, aiming to test, and at the same time, demonstrate the benefits and limitations of diffractive interface theory. 

\section{Applications of DIT}

\subsection{Dispersion of the guided waves}
\begin{figure}
\includegraphics[width=3in]{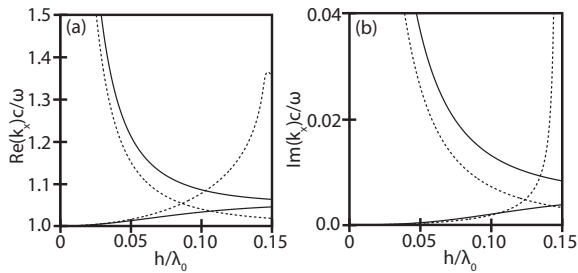}
\caption{Dispersion of guided waves supported by a thin metallic film as a function of film thickness $h$; solid and dashed lines represent analytical results and DIT calculations respectively }\label{fig:dispersion}
\end{figure}

We begin by calculating the dispersion of the guided modes supported by a thin homogeneous metallic film. The solutions of Maxwell's equation in this simple, yet illustrative, system are well known. These represent symmetric and anti-symmetric combinations of surface plasmon polaritons. As the film becomes thinner, the mode corresponding to symmetric distribution of the magnetic field approaches light line; while the mode that has an anti-symmetric distribution of magnetic field becomes increasingly confined to the surface of the film, with its modal index $n_{\rm eff}=c k_x/\omega \propto \lambda_0/h$\cite{raether13}. The behavior of these full-wave solutions of Maxwell's equations is illustrated in Fig.(\ref{fig:dispersion}) with solid lines. 

From the metasurface perspective, $\chi^{(g)}(x,y)=(\epsilon_m-1)/4\pi$. As can be explicitly verified, the matrices $\mathbb{K}_x$, $\mathbb{K}_y$, $\mathbb{X}_{xy}$, and $\mathbb{X}_{z}$ become diagonal, eliminating mixing between waves corresponding to different in-plane wavevectors. The terms corresponding to $\vec{k}_0$ can thus be isolated, and the solution to DIT can be found analytically. The dispersion of the two waves corresponding to the two guided waves supported by the film is shown in Fig.(\ref{fig:dispersion}) with dashed lines. It is clearly seen that in the limit as $h\ll\lambda_0$ DIT predictions converge to exact solutions of Maxwell's equations. For the parameters chosen here ($\lambda_0=1[\mu m],\epsilon^{(1)}=\epsilon^{(2)}=1,\epsilon^{(g)}=-10+1i$), DIT adequately describes the optical response of these modes when  $h\lesssim \lambda_0/20$.

\begin{figure}[b]
\includegraphics[width=3in]{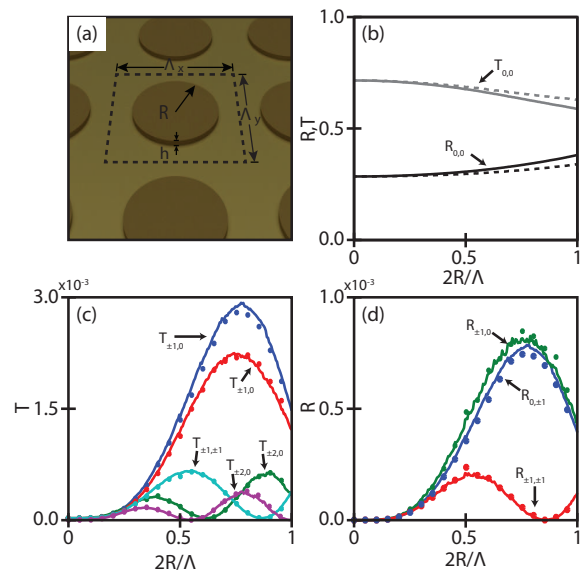}
\caption{(Color online) Diffraction by a metasurface formed by an array of plasmonic disks [$\epsilon^{(g)}=-10+1i$] deposited on dielectric substrate $[\epsilon^{(2)}=10.8]$; (a) schematic of the metasurface configuration; (b) 0th order reflection and transmission; solid and dashed lines represent RCWA and DIT calculations, respectively; (c),(d) higher-order reflection and transmission; RCWA (solid lines) and DIT(symbols) results}\label{fig:disks}
\end{figure}

\begin{figure*}[t]
\includegraphics[width=\textwidth]{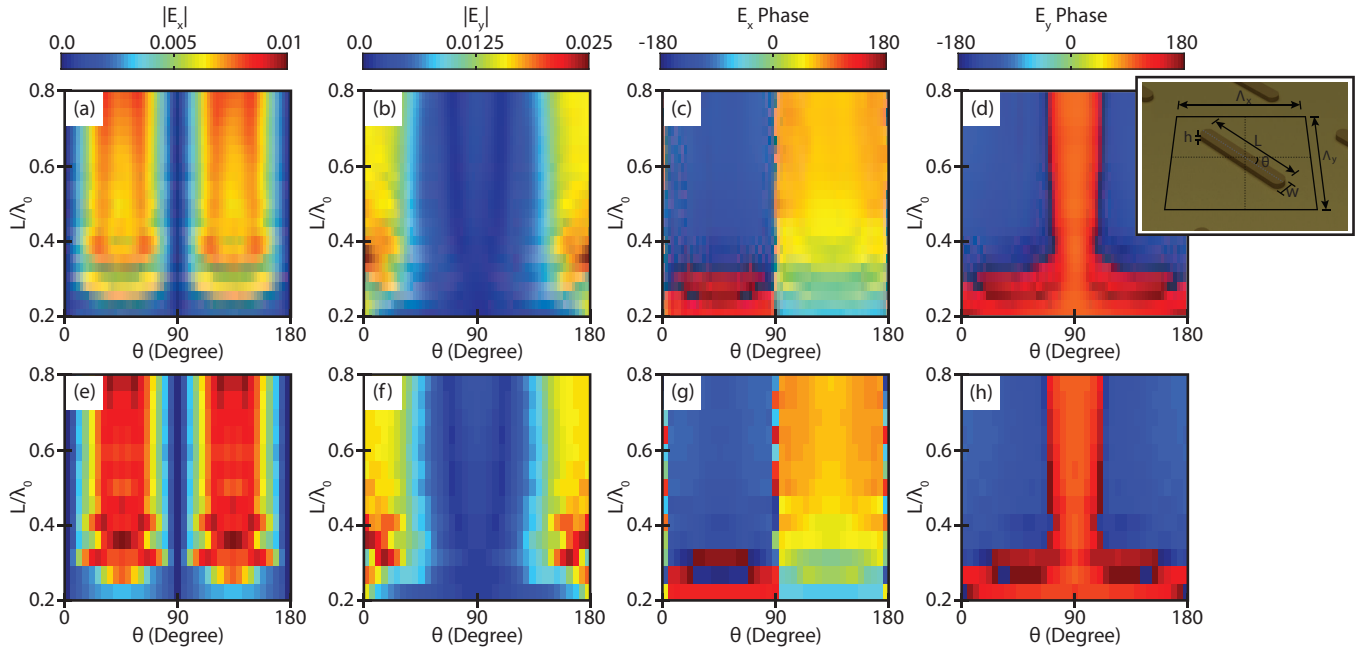}
\caption{(Color online) Amplitude  and phase of the electric field above the center of a highly metallic ($\epsilon^{(g)}=-2683+1367i$) nano-antenna array (see inset for geometry). DIT results shown in panels (a),(b),(c), and (d); RCWA results shown in panels (e),(f),(g), and (h); panels (a)\ldots(d) have four times as many pixels as panels (e)\ldots(h)}\label{fig:antenna}
\end{figure*}

\subsection{Generalized Refraction}
In order to verify the validity and efficiency of the developed technique for a main class of applications of metasurfaces, which involve generalized refraction and polarization control, we compare the results of DIT to the predictions of RCWA\cite{moharam81}. Since both techniques rely on mode expansion, the disagreement between the two essentially reflects the implication of approximating the metasurface as an optically thin layer. 

Techniques that rely on mode expansion require a certain number of modes in order to ensure convergence of the solution. In both DIT and RCWA, the number of required modes relates to the number of terms in discrete Fourier transform required to adequately represent the spatial profile of polarizability $\chi^{(g)}(x,y)$.  Our calculations indicate that DIT, presented in this work, converged as fast or faster than RCWA.  The real advantage of the above formalism is computational complexity.  The computational bottleneck for RCWA is the solution of an eigensystem [$\mathcal{O}(\frac{4}{3}n^3)$] that is required to calculate the propagation constants of the modes propagating through the periodic structure. At the same time, the limiting step of DIT is matrix division [$\mathcal{O}(\frac{2}{3}n^3)$]. The factor of two advantage can be dramatically improved with graphics processing unit (GPU) computing that strongly favors matrix division over eigenvalue problems\cite{GPU,cuBLAS}. In the end, our tests show that DIT runs almost an order of magnitude faster than RCWA for an equal number of modes, providing a compelling case for DIT for the design and optimization platform for optically thin metasurfaces. 

To analyze the validity of DIT, we first study the optical response of two classes of metasurfaces, (i) arrays of nano-disks on dielectric substrates and (ii) arrays of nano-antennae.  The behavior of the metasurfaces was studied in two different permittivity regimes, corresponding to highly metallic ($\epsilon=-2683+1367i$, permittivity of gold in the mid-IR\cite{johnson72}) and plasmonic ($\epsilon=-10+1i$, as achievable in the mid-IR with engineered highly-doped semi-conductors\cite{law14}) response of the metasurface material. In all calculations, the wavelength of excitation was fixed at $\lambda_0=8{\mu}m$ and the period was fixed at $\Lambda_x=\Lambda_y=15.92{\mu}m$.  

We first analyze the utility of DIT to calculate the diffraction efficiency of metasurfaces. To achieve this goal, the diffraction efficiency of a metasurface comprised of nano-disks were studied as a function of the disc radius.  The nano-disks are assumed to have a height of $h=\lambda_0/50$, and are deposited at the interface of a dielectric $[\epsilon^{(2)}=10.8]$ and air. The system is excited from the air side at normal incidence. The diffraction efficiencies of reflected and transmitted modes are summarized in Fig.\ref{fig:disks}. Overall, it is seen that DIT adequately represents the optical response of the structure. Minor deviations between DIT and RCWA predictions are seen at a high fill fraction.

Similar to FEM and FDTD calculations, DIT can be used to calculate not only diffraction efficiency, but also the field distribution across the system. However, in contrast to FEM, FDTD and similar techniques that calculate the {\it total field}, DIT is capable of analyzing the field distribution of {\it a particular wave} in the system. To illustrate this utility we analyze the response of the array of thin ($h=\lambda_0/100$) nano-antennae surrounded by air [Fig.\ref{fig:antenna}] for different lengths and angular orientation of the antennas and use the DIT to calculate the portion of the field corresponding to main reflection order right above the center of an antenna.  These results were then compared to RCWA and shown in Fig.\ref{fig:antenna}.  Once again, it is clearly seen that the DIT formalism adequately describes the field distribution in the system. Our calculations suggest a similar level of agreement in calculations of total field in this system and in the system where antennas are deposited at air-dielectric interface.  For a comparable number of $\{L,\theta\}$ combinations, DIT runs almost an order of magnitude faster than RCWA.

\begin{figure}[h]
\includegraphics[width=3in]{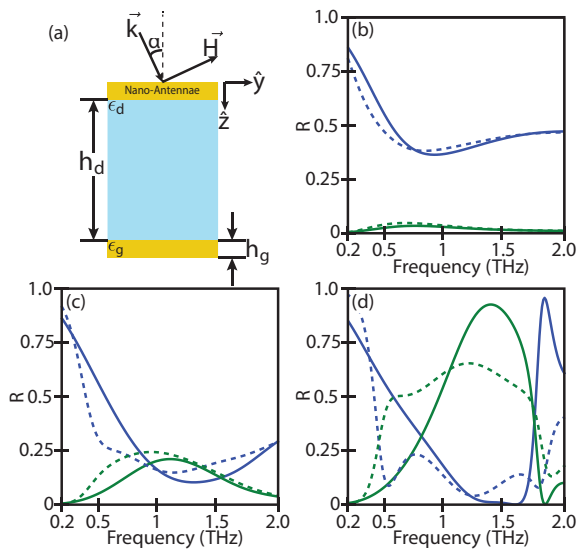}
\caption{(Color online) optical properties of nano-antenna-based polarization converter; (a) cross-section of the structure [see inset in Fig.\ref{fig:antenna}(d)  for geometry of the metasurface];   nano-antennae layer and gold ground layer are both 200nm thick $(h_g)$; periodicity $\Lambda_x=\Lambda_y=68{\mu}m$, antenna wire length $L=82{\mu}m$, antenna width $w=10{\mu}m$, dielectic layer height $h_d=33{\mu}m$ and angles $\theta=45^{\circ}$, and $\alpha=25^{\circ}$.  (b-d) Co-polarized (blue) and Cross-polarized (green) 0th order reflection calculated with DIT (solid lines) and RCWA (dashed lines) with (b) highly lossy spacer layer [$\epsilon_d=3(1+2i)$], (c) medium loss spacer layer [$\epsilon_d=3(1+.5i)$], and (d) low loss spacer layer [$\epsilon_d=3(1+.05i)$]}\label{fig:THz}.
\end{figure}

It is widely known that efficiency of metasurfaces for generalized refraction and other related applications is rather small. In practice, only $\lesssim 10\%$ of incident light is refracted into target diffraction order\cite{yu11,sun12,xiao13}. The efficiency of a metasurface, however, can be substantially enhanced when the metasurface is coupled to a homogeneous reflecting layer. In particular, antenna arrays incorporated into an optical stack with a highly reflective substrate have been recently suggested for applications in polarization conversion\cite{grady13}. In this structure, the polarization of incident beam gets rotated via {\it multiple} refections between metasurface and the perfectly conducting metallic plate below the surface [see Fig.\ref{fig:THz}]. 

While DIT correctly describes individual interaction of incident beam with a metasurface, the zero-thickness approximation introduces small errors into the optical response of the system. To understand the potential effect of accumulating of these small errors introduced by DIT, we consider the structure where the dielectric space between the metasurface and the PEC layer has variable loss (see Fig.\ref{fig:THz}a). In this geometry, increase of the loss leads to the decrease of the number of successive metasurface-metal reflections. Results of these studies are shown in Fig.\ref{fig:THz}(b-d). It is seen that when the number of successive reflections is relatively small, the predictions of DIT are almost identical to those of RCWA. However, as the number of interactions is increased, the accuracy of DIT is reduced. The number of successive reflections/transmissions through the metasurface serves as a limiting factor for DIT applications

\subsection{Optimizing the metasurface}
\begin{figure}[h]
\includegraphics[width=3in]{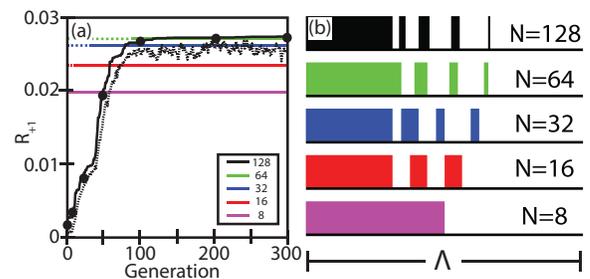}
\caption{ (Color online) Optimization of metasurface for maximum diffraction of light into $+1$ diffraction order; periodicity of the grating is fixed at $\Lambda=15.92{\mu}m$, binary mask width is fixed to $\Lambda/N$. Panel (a) represents the evolution of the sample used in Genetic Algorithm-based optimization routine, showing fitness ($R_{+1}$) for the best member of the population for a given generation (optimization step) [solid line] and mean fitness of the generation [dashed black line]. Black, green, blue, red, and magenta lines represent results for $N=128$,  $N=64$, $N=32$, $N=16$, and $N=8$ respectively; selected points of optimization procedure are validated with RCWA calculations (black dots).  Panel (b) shows results of the optimization, the unit cell that maximizes $R_{+1}$ for each of the chosen values $N$ after 300 optimization steps }\label{fig:Design}.
\end{figure}

Finally, we illustrate the potential of the proposed formalism in design and optimization of metasurfaces. For simplicity, here we optimize the parameters of 1D diffraction grating, maximizing the reflection into 1st diffraction order for a normal incident beam at $\lambda_0=8{\mu}m$. The unit cell of the grating is assumed to be represented as binary mask with fixed minimum width, $w=\Lambda/N$ (mimicking gratings fabricated with lighography-based approaches). This particular grating design naturally lends itself to a genetic optimization algorithm where the binary profile of the period serves as the chromosome of the member of the population, and the fitness function represents the percent of light diffracted into 1st order.  A highly metallic grating ($\epsilon=-2683+1367i$), with a height of $10nm$, on top of a dielectric substrate ($\epsilon=10.8$) was optimized for successively finer binary masks. The results of the optimization and optimization process are shown in Fig.\ref{fig:Design}.  The figure also compares the results of the DIT with RCWA solutions of Maxwell's equations for the purpose of an additional check of validity of the proposed technique. 

\section{conclusions}
To conclude, we presented a novel formalism to describe the interaction of light with optically thin diffractive systems (metasurfaces): diffractive interface theory. The formalism, which takes explicit advantage of the quasi-two-dimensional nature of metasurfaces, provides a direct link between the spatial profile of polarizability of metasurface and its diffraction (generalized refraction) properties. In our tests, GPU-based implementations of DIT have been demonstrated to run almost 10 times faster than comparable implementations of RCWA.  Applications of DIT for understanding the behavior of guided modes, calculations of generalized reflection of metasurfaces, calculations of field distributions in a system, and metasurface optimization have been demonstrated. The formalism can be straightforwardly extended to incorporate structures with anisotropic components, with strong magnetic response\cite{wong14}, as well as metasurfaces with non-planar geometries. 

This research is supported by NSF (grants ECCS-\#1102183 and DMR-\#1209761). 

\appendix
\section{Field profiles of the modes in the system}
As it can be directly verified, Eqs.(\ref{eq:curlE},\ref{eq:curlH}) admit harmonic $[\vec{E,H}\propto \exp(i k_x x+i k_y y)]$ solutions. In such solutions, the $z$-components of the fields can be related to the in-plane ($xy$) components via
\begin{align}
\mathcal{E}^{(b)}_{j,z}=&-\frac{c}{\omega\epsilon^{(b)}}\left(k_{j,x}\mathcal{H}^{(b)}_{j,y}-k_{j,y}\mathcal{H}^{(b)}_{j,x}\right)\\
\mathcal{H}^{(b)}_{j,z}=&\frac{c}{\omega}\left(k_{j,x}\mathcal{E}^{(b)}_{j,y}-k_{j,y}\mathcal{E}^{(b)}_{j,x}\right)
\nonumber
\end{align}

Straightforward substitution of the above equations into Eqs.(\ref{eq:curlE},\ref{eq:curlH}) demonstrates that the field of an individual mode from Eqs.(\ref{eq:Hamp}) can be decomposed into a linear combination of four plane waves with identical in-plane components of the wavevector ($k_{j,x} \hat{x}+k_{j,y} \hat{y}$). It is convenient to parameterize these waves according to their propagation [$+\hat{z}$ or $-\hat{z}$ direction] and polarization [transverse electric (TE) or transverse magnetic (TM), as defined with respect to their plane of incidence]. The linear relationship between the (harmonic) field profiles of the mode and the amplitudes of these four waves $\mathcal{C}^{(b)\pm}_p$ can be represented in matrix form as seen in Eq.(\ref{eq:fields}). The matrix $\mathbb{F}$ can be conveniently represented in block-matrix form as 

\begin{widetext}
\begin{equation}
\mathbb{F}^{(b)}=
\begin{pmatrix}
  -\mathbb{N}_{TE}\mathbb{K}_y \phi^{(b)+}& \mathbb{N}_{TM}\mathbb{K}_x \mathbb{K}_z^{(b)}/\epsilon^{(b)}\phi^{(b)+}&   -\mathbb{N}_{TE}\mathbb{K}_y \phi^{(b)-}	& \mathbb{N}_{TM}\mathbb{K}_x \mathbb{K}_z^{(b)}/\epsilon^{(b)}\phi^{(b)-} \\
  \mathbb{N}_{TE}\mathbb{K}_x\phi^{(b)+} & \mathbb{N}_{TM}\mathbb{K}_y \mathbb{K}_z^{(b)}/\epsilon^{(b)}\phi^{(b)+}&   \mathbb{N}_{TE}\mathbb{K}_x 	\phi^{(b)-}& \mathbb{N}_{TM}\mathbb{K}_y \mathbb{K}_z^{(b)}/\epsilon^{(b)} \phi^{(b)-}\\
  -\mathbb{N}_{TE}\mathbb{K}_x \mathbb{K}_z^{(b)}\phi^{(b)+}  	& -\mathbb{N}_{TM}\mathbb{K}_y\phi^{(b)+}_{TM}& \mathbb{N}_{TE}\mathbb{K}_x \mathbb{K}_z^{(b)}  \phi^{(b)-}	& \mathbb{N}_{TM}\mathbb{K}_y \phi^{(b)-} \\
  -\mathbb{N}_{TE}\mathbb{K}_x \mathbb{K}_z^{(b)}\phi^{(b)+}   	& \mathbb{N}_{TM}\mathbb{K}_x\phi^{(b)+}_{TM}& \mathbb{N}_{TE}\mathbb{K}_y \mathbb{K}_z^{(b)} \phi^{(b)-}	& -\mathbb{N}_{TM}\mathbb{K}_x\phi^{(b)-} \\ 
\end{pmatrix} 
\end{equation}
\end{widetext}
where matrices $\mathbb{K}_x$, $\mathbb{K}_y$, and $\mathbb{K}_z^{(b)}$ are defined in the text, and the elements of the remaining diagonal matrices are $(\phi^{(b)\pm})_{jj}=e^{\pm i k_{z,j}^{(b)} z}$, $(\mathbb{N}_{TE})_{jj}=\frac{1}{\sqrt{(k^2_{j,x}+k^2_{j,y})\left(k_{j,z}^{(b)^2}+1\right)}}$, and $(\mathbb{N}_{TM})_{jj}=\frac{1}{\sqrt{(k^2_{j,x}+k^2_{j,y})\left(k_{j,z}^{(b)^2}/\epsilon^{(b)}+1\right)}}$.

\section{Relationship between $\vec{E}$ and $\vec{P}$}
In the $x$ space, we define
$\vec{P}(x,y)=\chi(x,y)\vec{E}(x,y)$. In the Fourier space, this relationship becomes
\begin{eqnarray}
\sum_{m,n}\mathcal{P}_{m,n}e^{-i(k_{m,x}x+k_{n,y}y)}=\nonumber\\
\sum_{n,m}\sum_{p,q}\hat{\chi}_{m-p,n-q}\mathcal{E}_{p,q} e^{-i(k_{m,x}x+k_{n,y}y)}
\end{eqnarray}
leading to the relationship $\mathcal{P}_j=\sum_k\hat{\chi}_{j-k}\mathcal{E}_k$, once a single index is introduced to represent the ordered pairs as defined in the text. Equivalently, the last expression can be written in the matrix form as $\mathcal{P}=\tilde{\chi}\mathcal{E}$ .

\section{Derivation of Eq.(\ref{eq:MG})}

Direct application of a Fourier transform (similar to what is done above for $\vec{P}$ field) to Eqs.(\ref{eq:dH}), accompanied by substitution of Eq.(\ref{eq:fields}), results in the last two lines of Eq.(\ref{eq:MG}). 

Calculations of the discontinuity of electric field [Eqs.(\ref{eq:dE})], however, requires calculation of polarizability distribution. Here we use the fact that {\it microscopic} displacement field is continuous at each {\it microscopic} interface [substrate-grating, and grating-superstrate] of the system. This microscopic displacement field is related to electric field and polarization via $\vec{D}=\vec{E}+4 \pi\vec{P}$. Consequently, the discontinuity of electric field is written as 
\begin{equation}
\Delta \mathcal{E}_\alpha=-4\pi i h \mathbb{K}_\alpha \tilde{\chi}^{(b)}(\mathbb{I}+4\pi\tilde{\chi}^{(b)})^{-1}\mathcal{D}^{\rm avg}_z
\end{equation}
where  $D^{\rm avg}_\alpha(x,y)=[D_\alpha(x,y,z=+0)+D_\alpha(x,y,z=-0)]/2$ represents the average displacement field calculated inside bulk components of the metasurface.

\bibliography{DIT}

\end{document}